\documentclass[twocolumn]{elsart3p}
\usepackage{graphicx} 
\usepackage{dcolumn}  
\usepackage{longtable}
\usepackage{bm}       
\usepackage{subfigure}
\usepackage{subeqn}
\usepackage{boxedminipage}
\usepackage{slashed}
\begin{document}
\begin{frontmatter}
\title{Numerical Object Oriented Quantum Field Theory Calculations}
\author{M. Williams}
\address{Carnegie Mellon University, Pittsburgh PA, 15213}
\date{\today}

%
%
\begin{abstract} 
The {\tt qft++} package is a library of C++ classes that facilitate numerical
(not algebraic) 
quantum field theory calculations. Mathematical objects such as matrices,
tensors, Dirac spinors, polarization and orbital angular momentum tensors,
etc. are represented as C++ objects in {\tt qft++}. The package permits 
construction of code which closely resembles quantum field theory 
expressions, allowing for quick and reliable calculations. 
\end{abstract}

\end{frontmatter}
%
\section{\label{section:intro}Introduction}

It is often desirable to describe particle physics processes
using a covariant tensor formalism. This formalism contains Dirac spinors and
matrices, polarization and orbital angular momentum tensors, etc. 
For cases involving higher spin particles or large values of orbital 
angular momentum, the presence of high rank tensors can make explicit 
calculation of the desired quantities (typically scattering or decay 
amplitudes)
cumbersome at best and impossible at worst.

A common approach is to perform algebraic manipulations to reduce the 
number of tensor contractions prior to coding the expressions of interest.  
There are several software packages available to facilitate this approach, 
{\em e.g.} {\tt FeynCalc}~\cite{cite:feyncalc} and 
{\tt FeynArts}~\cite{cite:feynarts}. Packages like 
{\tt GRACE}~\cite{cite:grace} and {\tt CompHEP}~\cite{cite:comphep} 
take this a step further by (mostly automatically) producing distributions 
from physical models. While these packages are extremely useful in certain 
areas of physics ({\em e.g.} writing event generators), they are not ideal
for performing an event-based partial wave analysis (PWA).

In this type of analysis, numeric values for amplitudes must be calculated for
upwards of 100 million events (data and/or Monte Carlo). Thus, 
the amplitude calculation software must not only be easy to use, but also 
be capable of performing numerical calculations of expressions involving 
matrices and high-rank tensors using as little cpu time as possible.  
The {\tt qft++} package satisfies both of these criteria.

The operations of matrix multiplication and tensor contraction (the building
blocks of all such calculations) can
be broken down into nested loops; thus, they are easy to perform 
(numerically) on a computer. 
The {\tt qft++} package was developed to take advantage of this fact by
performing numerical calculations (not algebraic manipulations), 
making these types of calculations easier and more accessible to a larger 
portion of the physics community. 
The {\tt qft++} package has been used in a number of PWA's to date
(see, {\em e.g.},~\cite{cite:williams-thesis}) and has performed exceptionally
well.

One further point before discussing specific details about the software, this
package was designed to calculate tree-level expressions.  No work has been
done to provide a simple way of performing loop integrals.

\section{\label{section:overview}Overview}

The the main design goal for the application programming interface (API) was
to make the code resemble the mathematical expressions as closely as 
possible. To facilitate this goal, each type of mathematical object 
({\em e.g.} tensors, spinors, etc.) has a corresponding C++ object in
{\tt qft++}. Through the use of operator overloading, operations such as
matrix multiplication and tensor contraction are handled by the objects 
themselves; not by the user, {\em i.e.} the user never has to keep track of
any indices. 

As a very simple example, consider the electromagnetic vertex 
$e\bar{u}(p_1,m_1)\gamma^{\mu}u(p_2,m_2)\epsilon_{\mu}(p_{\gamma},m_{\gamma})$.
After variable declaration and initialization (discussed below), this 
expression can be written using {\tt qft++} as follows:
\\ \\
\begin{boxedminipage}{0.475\textwidth}
\begin{verbatim}
e*Bar(u1(m1))*gamma*u2(m2)*eps(mg);
\end{verbatim}
\end{boxedminipage}\\ \\ \\
The object types determine how ``multiplication'', {\em i.e.} the
{\tt *} operator, is to be performed. The matrix multiplications and tensor
contractions are handled internally by the objects.
This feature allows for self-documenting code, greatly reducing the probability
for mistakes. 
\section{\label{section:basic}Basic Operations}

The two main types of operations required in quantum field theory calculations,
but not contained in standard C++, involve matrices and tensors. 
From these constituents, it is easy to build classes which handle Dirac 
matrices, covariant projection operators, etc. This section describes how
the {\tt qft++} package implements these basic types of operations.

This package makes heavy use of template classes; thus, a number of template
utilities have been developed for performance and/or API reasons. Among these
are compile-time detection of inheritance and parameter passing 
optimization~\cite{cite:alexandrescu}, along with selective inclusion of class
methods~\cite{cite:boost}.

Unlike many standard C++ template classes,
all of the template classes defined in the {\tt qft++} package are designed
such that instantiations of different types are fully compatible, provided the
types themselves have the necessary operators defined, 
{\em e.g.} the following code is legal:
\\ \\
\begin{boxedminipage}{0.475\textwidth}
\begin{verbatim}
SomeClass<T1> sc1; 
SomeClass<T2> sc2; 
sc1*sc2;
\end{verbatim}
\end{boxedminipage}\\ \\ \\
if the operator {\tt *} is defined between types {\tt T1} and {\tt T2}.

\subsection{\label{section:basic:matrix}Matrix Operations}

Matrix operations are handled by the template class \texttt{Matrix}, which 
can store any data type which can be stored in a C++ STL container class,
{\em i.e.} it can store any object which can be stored by
\texttt{std::vector}. A complete set of methods are provided including those
useful in Quantum Field theory calculations, such as {\tt Trace} and 
{\tt Adjoint}, along with all the necessary operators.

\subsection{\label{section:basic:tensor}Tensor Operations}

Tensor operations are handled by the template class {\tt Tensor}, which can
also store any data type which can be stored in a C++ STL container class;
however, the most useful types are typically {\tt double} and 
{\tt complex<double>}. 
Methods are provided to perform Lorentz transformations, symmetrization
and a number of other functions.

A number of operators are provided to perform tensor
contractions.
The {\tt Tensor} class assumes that all indices are either raised or 
lowered.  The field theory objects discussed in Section~\ref{section:objects}
use the former, {\em i.e.} they are contravariant.    
The {\tt *} operator performs standard multiplication if either object is not 
a {\tt Tensor} and contraction of a single index otherwise.  For example, if
{\tt x} is a rank-2 {\tt Tensor}, then the code snippet {\tt 3*x} simply 
multiplies each of the elements of {\tt x} by three; however, the snippet
{\tt x*y} will evaluate the expression $x^{\mu\nu} y_{\nu}{}^{\rho}$ if {\tt y}
is also a rank-2 {\tt Tensor}.  Full contraction of all possible indices is
performed using the {\tt |} operator; thus, {\tt x|y} evaluates the 
expression $x^{\mu\nu} y_{\mu\nu}$. 

\begin{table}[h!]
\begin{center}
\begin{tabular}{c|c} 
  \qquad Symbolic Expression \hspace{0.025\textwidth} 
  & \qquad \qquad {\tt qft++} Code \hspace{0.05\textwidth} 
  \\\hline
  $x^{\mu} y_{\mu}$ & {\tt x*y} \\ 
  $x^{\mu} y^{\nu}$ & {\tt x\%y} \\ 
  $x^{\mu_1\mu_2\ldots\mu_n} y_{\mu_1\mu_2\ldots\mu_n}$ & {\tt (x|y)} \\ \hline
\end{tabular}
\\
\vspace{0.01\textheight}
\caption[]{\label{table:tensor:ops} 
  Example tensor operations using the {\tt Tensor} template class.}
\end{center}
\end{table}

Additional classes are provided for the Minkowski metric, $g_{\mu\nu}$
({\tt MetricTensor}), and the Levi-Civita tensor, 
$\epsilon_{\mu\nu\alpha\beta}$ ({\tt LeviCivitaTensor}). The 
{\tt Vector4} template class, which is derived from {\tt Tensor}, 
provides a number of additional methods specific to 4-vectors, 
{\em e.g.} {\tt CosTheta} and {\tt Beta}.
\section{\label{section:objects}Object-Oriented Field Theory}

From the {\tt Tensor} and {\tt Matrix} classes discussed above, all of the 
necessary objects for quantum field theory calculations can be constructed.
In this section, a brief overview of the formalism will be presented followed
by a discussion of the corresponding {\tt qft++} class.
More detailed descriptions of the formalism are given 
in~\cite{cite:rarita,cite:zemach-form}. Recent examples of applying this 
formalism to PWA - for which this code was developed - can be found
in~\cite{cite:williams-thesis,cite:chung,cite:anisovich}.

\subsection{\label{section:integral}Integral Spin Wave Functions}

The wave function of a particle with integral spin-$J$, 4-momentum $p$ and 
spin projection to some quantization axis $m$, is 
described by a rank-$J$ tensor, $\epsilon_{\mu_1\ldots\mu_J}(p,m)$. 
The Rarita-Schwinger conditions for integral spin-$J$ are
\begin{subequations}
  \label{eq:integral-spin-rs}
  \begin{equation}
    p^{\mu_i}\epsilon_{\mu_1\mu_2\ldots\mu_i\ldots\mu_J}(p,m) = 0  
  \end{equation}
  \begin{equation}
    \epsilon_{\mu_1\ldots\mu_i\ldots\mu_j\ldots\mu_J}(p,m) = 
    \epsilon_{\mu_1\ldots\mu_j\ldots\mu_i\ldots\mu_J}(p,m) 
  \end{equation}
  \begin{equation}
    g^{\mu_i\mu_j}\epsilon_{\mu_1\mu_2\ldots\mu_i\ldots\mu_j\ldots\mu_J}(p,m)
     =  0,
  \end{equation}
\end{subequations}
for any $\mu_i,\mu_j$, and reduce the number of independent elements from $4^J$
to $(2J+1)$. 

The spin-$J$ projection operator, defined as
\begin{eqnarray}
  \label{eq:integral-spin-proj}
  P^{(J)}_{\mu_1\mu_2\ldots\mu_J\nu_1\nu_2\ldots\nu_J}(p) 
  \hspace{0.25\textwidth}\nonumber \\
  \hspace{0.05\textwidth} = \sum\limits_{m}
  \epsilon_{\mu_1\mu_2\ldots\mu_J}(p,m)\epsilon^*_{\nu_1\nu_2\ldots\nu_J}(p,m),
\end{eqnarray}
is used to construct the particle's propagator.

As a simple example, consider a massive spin-1 particle.
The Rarita-Schwinger conditions for spin-1 simply require
$p^{\mu}\epsilon_{\mu}(p,m) = 0$. Thus, in the particle's rest frame the energy
component of the wave function is zero. The spatial components are then chosen
to be
\begin{equation}
\vec{\epsilon}(\pm 1) = \mp \frac{1}{\sqrt{2}}(1,\pm i,0),\hspace{0.2in} 
  \vec{\epsilon}(0) = (0,0,1).
\end{equation}
The wave function can be obtained in any other frame through the use of 
Lorentz transformations.
The spin-1 projection operator is given by
\begin{equation}
  \label{eq:spin1-proj}
  P^{(1)}_{\mu\nu}(p) = \sum\limits_{m} 
  \epsilon_{\mu}(p,m)\epsilon^{*}_{\nu}(p,m) 
  = -g_{\mu\nu} + \frac{p_{\mu}p_{\nu}}{w^2},
\end{equation}
where $w$ is the mass of the particle. 

Wave functions for particles with integral spin are handled in {\tt qft++} by
the {\tt PolVector} class. The spin is set using the constructor, {\em e.g}
{\tt PolVector eps(3)} would be used for a spin-3 particle. The {\tt PolVector}
object must then be initialized for a given 4-momentum. At this point, the user
can decide whether or not the particle is to be on-shell. For example, if a
spin-1 particle is initialized on-shell then the projection operator is given
by
\begin{equation}
  P^{(1)}_{\mu\nu}(p)
  = -g_{\mu\nu} + \frac{p_{\mu}p_{\nu}}{p^2},
\end{equation}
otherwise, it is calculated using (\ref{eq:spin1-proj}). This option is also
available in {\tt qft++} for half-integral spin particles.

After initialization,
the sub-states can be accessed via calls like {\tt eps(m)}, which returns the
{\tt Tensor<complex<double> >} object for sub-state $m$. The spin-$J$ projection
operator can be accessed easily in the code using the method 
{\tt eps.Projector()}, which returns an object of type 
{\tt Tensor<complex<double> >} with a rank of $2J$ .

\begin{table*}[]
\begin{center}
\begin{tabular}{c|c|c} 
  {\tt qft++} Class & Symbol & Concept \\ \hline
  {\tt Matrix<T>} & $a_{ij}$ & matrices of any dimension \\ \hline
  {\tt Tensor<T>} & $x_{\mu_1\ldots\mu_n}$ & tensors of any rank \\ 
  {\tt MetricTensor} & $g_{\mu\nu}$ & Minkowski metric \\ 
  \hspace{0.05\textwidth}
  {\tt LeviCivitaTensor} 
  \hspace{0.05\textwidth}
  & 
  \hspace{0.05\textwidth}
  $\epsilon_{\mu\nu\alpha\beta}$ 
  \hspace{0.05\textwidth}
  & 
  \hspace{0.05\textwidth}
  totally 
  anti-symmetric Levi-Civita tensor 
  \hspace{0.05\textwidth}
  \\ \hline
  {\tt DiracSpinor} & $u_{\mu_1\ldots\mu_{J-1/2}}(p,m)$ & half-integral spin
  wave functions \\ 
  {\tt DiracAntiSpinor} & $v(p,m)$ & spin-1/2 anti-particle wave functions 
  \\ \hline
  {\tt DiracGamma} & $\gamma^{\mu}$ & {} \\ 
  {\tt DiracGamma5} & $\gamma^5$ & Dirac matrices \\ 
  {\tt DiracSigma} & $\sigma^{\mu\nu}$ & {} \\ 
  \hline  
  {\tt PolVector} & $\epsilon_{\mu_1\ldots\mu_J}(p,m)$ & integral spin wave
  functions \\ \hline
  {\tt OrbitalTensor} & $L^{(\ell)}_{\mu_1\ldots\mu_{\ell}}$ & 
  orbital angular momentum tensors \\ \hline
  
\end{tabular}
\\
\vspace{0.01\textheight}
\caption[]{\label{table:qft-classes} 
  A partial list of {\tt qft++} classes, a complete list can be found 
  at~\cite{cite:qft++website}.
}
\end{center}
\end{table*}

\subsection{\label{section:half-integral}Half-Integral Spin Wave Functions}

The wave function for a spin-1/2 particle is described by a 4-component Dirac 
spinor, denoted as $u(p,m)$, where $p$ and $m$ again represent the 4-momentum 
and spin projection of the particle. The representation chosen here leads to
the following form of the spinors:
\begin{equation}
  u(p,m)  = \sqrt{E+w}
  \left( \begin{array}{c} \chi(m) \\ \frac{\vec{\sigma}\cdot\vec{p}}{E + w} 
    \chi(m) \end{array}  \right),
\end{equation}
where $E(w)$ is the energy(mass) of the particle and
$\chi(m)$
are the standard non-relativistic 2-component spinors.

Wave functions for particles with higher half-integral spin,
denoted as $u_{\mu_1\ldots\mu_{J-1/2}}(p,m)$, are constructed using
tensor products of integral spin wave functions and the spin-1/2 spinors 
described above. The Rarita-Schwinger conditions for half-integral spin-$J$ 
are~\cite{cite:rarita}
\begin{subequations}
  \label{eq:half-integral-spin-rs}
  \begin{equation}
    (\gamma^{\mu}p_{\mu}-w) u_{\mu_1\ldots\ldots\mu_{J-1/2}}(p,m) = 0 
  \end{equation}
  \begin{eqnarray}
    u_{\mu_1\ldots\mu_i\ldots\mu_j\ldots\mu_{J-1/2}}(p,m)  
    \hspace{0.15\textwidth}\nonumber\\
    \hspace{0.15\textwidth}
    = u_{\mu_1\ldots\mu_j\ldots\mu_i\ldots\mu_{J-1/2}}(p,m) 
  \end{eqnarray}
  \begin{equation}
    p^{\mu_i}u_{\mu_1\ldots\mu_i\ldots\mu_{J-1/2}}(p,m) = 0 
  \end{equation}
  \begin{equation}
    \gamma^{\mu_i}u_{\mu_1\ldots\mu_i\ldots\mu_{J-1/2}}(p,m) = 0 
  \end{equation}
  \begin{equation}
    g^{\mu_i\mu_j}u_{\mu_1\ldots\mu_i\ldots\mu_j\ldots\mu_{J-1/2}}(p,m) = 0,
  \end{equation}
\end{subequations}
for any $\mu_i,\mu_j$, 
and reduce the number of independent elements from $4^{J+1/2}$ to $(2J+1)$.

The spin-$J$ projection operator is then defined as
\begin{eqnarray}
  \label{eq:half-integral-spin-proj}
  P^{(J)}_{\mu_1\ldots\mu_{J-1/2}\nu_1\ldots\nu_{J-1/2}}(p) 
  = \frac{1}{2w} \hspace{0.15\textwidth}
  \nonumber \\
  \hspace{0.04\textwidth}
  \times \sum\limits_{m} u_{\mu_1\ldots\ldots\mu_{J-1/2}}(p,m) 
  \bar{u}_{\nu_1\ldots\ldots\nu_{J-1/2}}(p,m).
\end{eqnarray}
For example, the spin-1/2 projection operator is simply
$P^{(1/2)}(p) = \frac{1}{2w}(p^{\mu}\gamma_{\mu} + w)$,
while for spin-3/2 the projection operator is given by
\begin{eqnarray}
  P^{(\frac{3}{2})}_{\mu\nu}(p) =  - P^{(\frac{1}{2})}(p) 
  \hspace{0.25\textwidth}\nonumber \\
  \hspace{0.05\textwidth}
  \times \left(
  P^{(1)}_{\mu\nu}(p) + \frac{1}{3}P^{(1)}_{\mu\alpha}(p)\gamma^{\alpha}
  P^{(1)}_{\nu\beta}(p)\gamma^{\beta}\right).
\end{eqnarray}

Wave functions for particles with half-integral spin are handled in 
{\tt qft++} by the {\tt DiracSpinor} class. The spin is set using the
constructor, {\em e.g.} {\tt DiracSpinor u(3/2.)} would be used for a 
spin-3/2 particle. As with the {\tt PolVector} class, the {\tt DiracSpinor}
class must be initialized for a given 4-momentum. The sub-states can then be
accessed via {\tt u(m)}, which returns the 
{\tt Matrix<Tensor<complex<double> > >} object for sub-state $m$. 
The spin-$J$ projection operators can be easily accessed using the method
{\tt u.Projector()} which returns a $4\times4$ {\tt Matrix} of 
rank-$(2J-1)$ {\tt Tensor<complex<double> >} objects. We also note here that
the quantity $\bar{u}(p,m)$ is obtained using the function {\tt Bar(u(m))}.

\subsection{\label{section:dirac-gamma}Dirac Matrices}

Classes are also provided to handle the Dirac matrices, $\gamma^{\mu}$
({\tt DiracGamma}) and $\gamma^5$ ({\tt DiracGamma5}), along with 
$\sigma^{\mu\nu} \equiv \frac{i}{2}[\gamma^{\mu},\gamma^{\nu}]$
({\tt DiracSigma}). 
Each of these classes is derived from the common base class
{\tt Matrix<Tensor<complex<double> > >}. Thus, they inherit all of the 
necessary {\tt Matrix} and {\tt Tensor} operators.

\subsection{\label{section:orbital}Orbital Angular Momentum Tensors}

Two particles, with 4-momenta $p_a$ and $p_b$, can be coupled to a state of 
pure orbital angular momentum, $\ell$, using the operators
$L^{(\ell)}_{\mu_1\mu_2\ldots\mu_{\ell}}$.
The total and relative momenta are defined as $P = p_a + p_b$ and 
$p_{ab} = \frac{1}{2}(p_a - p_b)$ respectively. The orbital-angular-momentum
operators are then built using the relative momentum and the spin-$\ell$ 
projection operator as follows:
\begin{equation}
  \label{eq:orbital-ops-def}
  L^{(\ell)}_{\mu_1\mu_2\ldots\mu_{\ell}} \propto
  P^{(\ell)}_{\mu_1\mu_2\ldots\mu_{\ell}\nu_1\nu_2\ldots\nu_{\ell}}(P)
  p_{ab}^{\nu_1}p_{ab}^{\nu_2}\ldots p_{ab}^{\nu_{\ell}}.
\end{equation}
These operators satisfy the Rarita-Schwinger conditions
\begin{subequations}
  \label{eq:orbital-ops-rs}
  \begin{equation}
    P^{\mu_i}L^{(\ell)}_{\mu_1\mu_2\ldots\mu_i\ldots\mu_{\ell}} =  0 
  \end{equation}  
  \begin{equation}
    L^{(\ell)}_{\mu_1\mu_2\ldots\mu_i\ldots\mu_j\ldots\mu_{\ell}}  = 
    L^{(\ell)}_{\mu_1\mu_2\ldots\mu_j\ldots\mu_i\ldots\mu_{\ell}} 
  \end{equation}
  \begin{equation}
    g^{\mu_i\mu_j}
    L^{(\ell)}_{\mu_1\mu_2\ldots\mu_i\ldots\mu_j\ldots\mu_{\ell}} =  0,
  \end{equation}
\end{subequations}
for any $\mu_i,\mu_j$, which insure that they have $(2\ell+1)$ independent 
elements. 

In the {\tt qft++} package, orbital-angular-momentum operators are handled by
the {\tt OrbitalTensor} class. This class inherits from {\tt Tensor<double>};
thus, after construction it can be used just like any other tensor. Setting the
tensor elements of $L^{(\ell)}_{\mu_1\ldots\mu_{\ell}}$ ({\tt OrbitalTensor} 
object {\tt orbL}) for 4-momenta $p_a$ and $p_b$ ({\tt Vector4<double>} objects
{\tt pa} and {\tt pb}) is done by simply calling
{\tt orbL.SetP4(pa,pb)}.

\subsection{\label{section:utils}Additional Utilities}

Functions are also provided to calculate useful quantities such as 
Clebsch-Gordon coefficients, Wigner D-functions, Breit-Wigner and Regge 
propagators, etc. 
\section{\label{section:examples}Example Applications}

In this section, a few simple examples will be examined. The {\tt qft++}
package computes the values of expressions numerically; thus, the 4-momenta
of the particles involved must be known. In an event-based partial wave 
analysis, these would be obtained from the experimental data and/or Monte 
Carlo events. In the case that one wants to calculate theoretical angular
distributions, cross sections, polarization observables, etc., events with 
the desired kinematics must be generated as input for the {\tt qft++} code.
In the examples below, the assumption is made that one of these methods is 
employed. For
the example plots shown in this section, the latter method was used.

\subsection{\label{section:examples:jpsi}
  $X(2^-) \rightarrow \omega K \rightarrow \pi^+\pi^-\pi^0K$}

Consider the decay of a particle, $X$, with spin-parity $J^P = 2^-$ into an
$\omega$ and $K$ via $F$-wave.
The invariant decay amplitude for this process is proportional to
\begin{equation}
  \label{eq:spin2-amp}
  \mathcal{A} \propto \epsilon^{*}_{\mu}(p_{\omega},m_{\omega}) 
  L^{(3)\mu\nu\alpha}(p_{\omega K}) \epsilon_{\nu\alpha}(P,M),
\end{equation}
where $p_{\omega},m_{\omega}(P,M)$ are the momentum and spin projection of the 
$\omega(X)$ and 
$p_{\omega K}$ is the relative momentum of the $\omega K$ system.

For this example, assume that the $X$ is produced via $e^+e^-$ annihilation
resulting in population of only the $M = \pm1$ sub-states.
The decay distribution is then obtained by calculating the intensity
\begin{equation}
  \label{eq:spin2-int}
  \mathcal{I} \propto \sum\limits_{M=\pm1}\sum\limits_{m_{\omega}=\pm1,0} 
  |\mathcal{A}|^2.
\end{equation}

To calculate this distribution using {\tt qft++}, the necessary variables  must
first be declared:\\ \\
\begin{boxedminipage}{0.475\textwidth}
\begin{verbatim}
PolVector epso; // omega
PolVector epsx(2); // X
OrbitalTensor orb3(3); // L^3
Tensor<complex<double> > amp; 
Vector4<double> p4o,p4k,p4x;
\end{verbatim}
\end{boxedminipage}\\ \\ \\
For each point at which $\mathcal{I}$ is to be calculated, the 4-momenta 
must be set and the polarization states initialized using calls to 
{\tt SetP4}; however, if the kinematics are such that the $X$ mass is
constant, then {\tt p4x} and {\tt epsx} will only need to be initialized once.

In the code, a loop would then be performed over all values of $\cos{\theta}$
(the decay angle of the $\omega$ in the $X$ rest frame) for which the decay
intensity is to be calculated (or over events). 
At each point, the calculation would be 
performed as: \\ \\
\begin{boxedminipage}{0.475\textwidth}
\begin{verbatim}
double intensity = 0.;
for(Spin m = -1; m <= 1; m+=2){ 
    for(Spin mo = -1; mo <= 1; mo++){ 
        amp = conj(epso(mo))*orb3|epsx(m);
        intensity += norm(amp());
    }
}
\end{verbatim}
\end{boxedminipage}\\ \\ \\
In this way, the value of $\mathcal{I}$ can be calculated at any number of
points in $\cos{\theta}$ (or, for any number of events).

It is worth examining this more carefully.  The tensor contractions in the code
above are performed by first evaluating the {\tt *} operator which contracts 
$\epsilon^{*}_{\mu}(p_{\omega},m_{\omega})$ into the first index of 
$ L^{(3)\mu\nu\alpha}(p_{\omega K})$.
The result is a rank-2 {\tt Tensor} whose two indices are contracted
via the {\tt |} operator into both indices of 
$\epsilon_{\nu\alpha}(P,M)$.  
The result is a rank-0 {\tt Tensor<complex<double> >} whose value is accessed
via the {\tt ()} operator.

To check that the code is working for this simple example, the intensity in 
(\ref{eq:spin2-int}) can be calculated in the $X$ rest frame by making a slight
modification to the non-relativistic helicity formalism solution as
follows:
\begin{equation}
  \label{eq:spin2-hel-def}
  \mathcal{I} \propto \sum\limits_{M,\lambda}
  |f(E_{\omega}/w_{\omega},\lambda)
  (301\lambda|2\lambda) d^2_{M\lambda}(\theta)|^2,
\end{equation}
where $E_{\omega},w_{\omega},\lambda$ are the energy, mass and helicity of
the $\omega$ and
$f(x,\pm 1) = 1$, $f(x,0) = x$ accounts for the effects of the boosts on
the covariant $\omega$ helicity states. Notice that as 
$E_{\omega} \rightarrow w_{\omega}$, the non-relativistic solution is 
recovered.

\begin{figure}
  \begin{center}
  \includegraphics[width=0.5\textwidth]{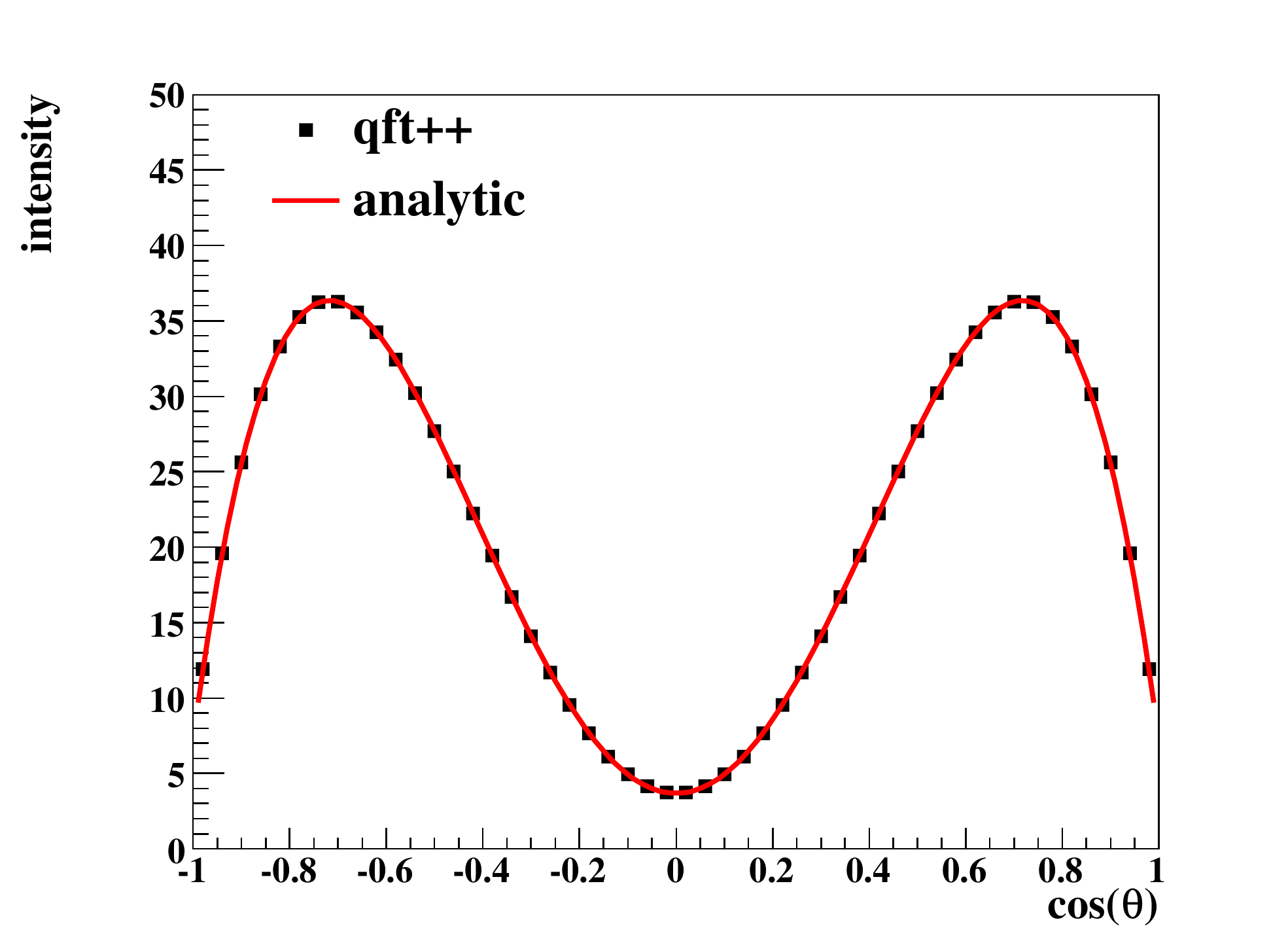}
  \caption[]{\label{fig:spin2-dist}
    (Color Online)
    Intensity vs $\cos{\theta}$: Angular distribution 
    calculated for the decay 
    ${X(2^-) \rightarrow \omega K}$ using 
    {\tt qft++} (black filled squares) compared with an analytic solution
    (red line). See text for details.
  }
  \end{center}
\end{figure}

The expression in
(\ref{eq:spin2-hel-def}) can then be rewritten purely in terms of the decay
angle as follows:
\begin{eqnarray}
  \label{eq:spin2-hel}
  \mathcal{I} \propto (2\cos^2{\theta} - 1)^2 + \cos^2{\theta} 
  \hspace{0.15\textwidth}\nonumber \\
  \hspace{0.2\textwidth}
  + 9\left(\frac{E_{\omega}}{w_{\omega}}\right)^2 \sin^2{\theta}\cos^2{\theta}.
\end{eqnarray}
Figure~\ref{fig:spin2-dist} shows the angular distributions obtained by 
calculating (\ref{eq:spin2-amp}) using {\tt qft++} compared to the 
modified helicity formalism expression given in (\ref{eq:spin2-hel}),
normalized to have the same integral as the {\tt qft++} solution. 
The two calculations give the same angular distributions, {\em i.e.} the code
is working properly.

This example can be extended by considering the secondary decay 
${\omega \rightarrow \pi^+ \pi^- \pi^0}$. The amplitude for this process
can be written as
\begin{equation}
  \label{eq:omega-decay-amp}
  \mathcal{A}_{\omega \rightarrow \pi^+ \pi^- \pi^0} \propto 
  i \epsilon_{\mu\nu\alpha\beta}p_{\pi^+}^{\nu}p_{\pi^-}^{\alpha}
  p_{\pi^0}^{\beta} \epsilon^{\mu}(p_{\omega},m_{\omega}),
\end{equation}
which, in the $\omega$ rest frame, simplifies to
\begin{equation}
    \mathcal{A}^{m_{\omega}}_{\omega \rightarrow \pi^+ \pi^- \pi^0} \propto
    i \left( \vec{p}_{\pi^+} \times \vec{p}_{\pi^-} \right)
    \cdot  \vec{\epsilon}(m_{\omega}).
\end{equation}
This is the standard non-relativistic result~\cite{cite:zemach}.

To incorporate this decay, (\ref{eq:spin2-amp}) must be rewritten as
\begin{equation}
  \label{eq:spin2-amp-with-decay}
  \mathcal{A} \propto \omega_{\mu}(p_{\omega}) 
  L^{(3)\mu\nu\alpha}(p_{\omega K}) \epsilon_{\nu\alpha}(P,M),
\end{equation}
where,
\begin{equation}
  \omega_{\mu}(p_{\omega}) \propto 
  \frac{i P^{(1)\mu\nu}(p_{\omega})}{p_{\omega}^2 - w_{\omega}^2 + i w_{\omega}\Gamma_{\omega}}
  \epsilon_{\nu\rho\alpha\beta}p_{\pi^+}^{\rho}p_{\pi^-}^{\alpha}
  p_{\pi^0}^{\beta},
\end{equation}
and $w_{\omega},\Gamma_{\omega}$ are the mass and width of the $\omega$.

To add this decay to the {\tt qft++} calculation, the following variables need 
to be defined: \\ \\
\begin{boxedminipage}{0.475\textwidth}
\begin{verbatim}
complex<double> i(0,1);
Tensor<complex<double> > omega;
LeviCivitaTensor levi;
Vector4<double> p4pip,p4pim,p4pi0;
\end{verbatim}
\end{boxedminipage}\\ \\ \\
Then, for each point the intensity is calculated as:
\\ \\
\begin{boxedminipage}{0.475\textwidth}
\begin{verbatim}
omega = epso.Projector()*levi
    *p4pip*p4pim*p4pi0
    *BreitWigner(p4o,0.78256,0.00844);
double intensity = 0.;
for(Spin m = -1; m <= 1; m+=2){ 
    amp = omega*orb3|epsx(m);
    intensity += norm(amp());
}
\end{verbatim}
\end{boxedminipage}\\ \\ \\
Since all of the objects are covariant, no extra boosts or rotations to the 
$\omega$ rest frame are required.

\subsection{\label{section:examples:delta}
  $\pi p \rightarrow \Delta \rightarrow \pi p$}

As an example involving half-integral spin particles, consider the reaction
${\pi p \rightarrow \Delta(1232) \rightarrow \pi p}$. The invariant scattering
amplitude for this process is proportional to
\begin{equation}
  \label{eq:delta-amp}
  \mathcal{M} \propto \bar{u}(p_f,m_f) p_f^{\mu} P^{(\frac{3}{2})}_{\mu\nu}(P)
  p_i^{\nu} u(p_i,m_i),
\end{equation}
where $p_i(p_f)$ and $m_i(m_f)$ are the 4-momentum and
canonical spin projection of the initial(final) proton respectively. 
For this example, the mass-dependence of the propagator will be ignored. In the
code, adding this would simply involve multiplying the amplitude by a 
{\tt complex<double>}.

The distribution in the scattering angle, {\em i.e.} the angle between the 
initial and final protons 
(${\cos{\theta} = \hat{p}_i\cdot\hat{p}_f}$),
is obtained by calculating the scattering intensity
\begin{equation}
  \mathcal{I} \propto \sum\limits_{m_i,m_f} |\mathcal{M}|^2.
\end{equation}

To calculate this distribution using {\tt qft++}, the following variables must
be declared:\\ \\
\begin{boxedminipage}{0.475\textwidth}
\begin{verbatim}
DiracSpinor ui,uf; // proton spinors
DiracSpinor delta(3/2.); 
Matrix<complex<double> > amp(1,1); 
Vector4<double> p4_i,p4_f,p4_delta;
\end{verbatim}
\end{boxedminipage}\\ \\ \\
For each point at which $\mathcal{I}$ is to be calculated, the 4-momenta 
must be set and the spinors initialized using calls to {\tt SetP4}; however,
if the 4-momentum of any given particle does change from point to point, then
its corresponding object would only need to be initialized once.

In the code, a loop would then be performed over all values of $\cos{\theta}$
for which the scattering intensity is to be calculated (or, again, over all 
events). At each point, the
calculation would be performed as:\\ \\
\begin{boxedminipage}{0.475\textwidth}
\begin{verbatim}
double intensity = 0.;
for(Spin m_i = -1/2.; m_i <= 1/2.; m_i++){
    for(Spin m_f = -1/2.; m_f <= 1/2.; m_f++){
        amp = Bar(uf(m_f))*p4_f
                   *delta.Projector()*p4_i*ui(m_i);
        intensity += norm(amp(0,0));
    }
}
\end{verbatim}
\end{boxedminipage}\\ \\ \\
In this way, the value of $\mathcal{I}$ can be calculated at any number of
points in $\cos{\theta}$. 

If the $\Delta(1232)$ projector is on-shell (see discussion in 
Section~\ref{section:integral}), then the numerical calculations can be checked
in the overall center-of-mass frame using the non-relativistic helicity 
formalism. In this frame, the scattering intensity is proportional to
\begin{equation}
  \label{eq:delta-hel}
  \mathcal{I} \propto \sum\limits_{\lambda_i,\lambda_f}
  |d^{\frac{3}{2}}_{\lambda_i\lambda_f}(\theta)|^2 
  \propto 1 + 3\cos^2{\theta},
\end{equation}
where $\lambda_i(\lambda_f)$ are the initial(final) proton helicities. 

\begin{figure}
  \begin{center}
  \includegraphics[width=0.5\textwidth]{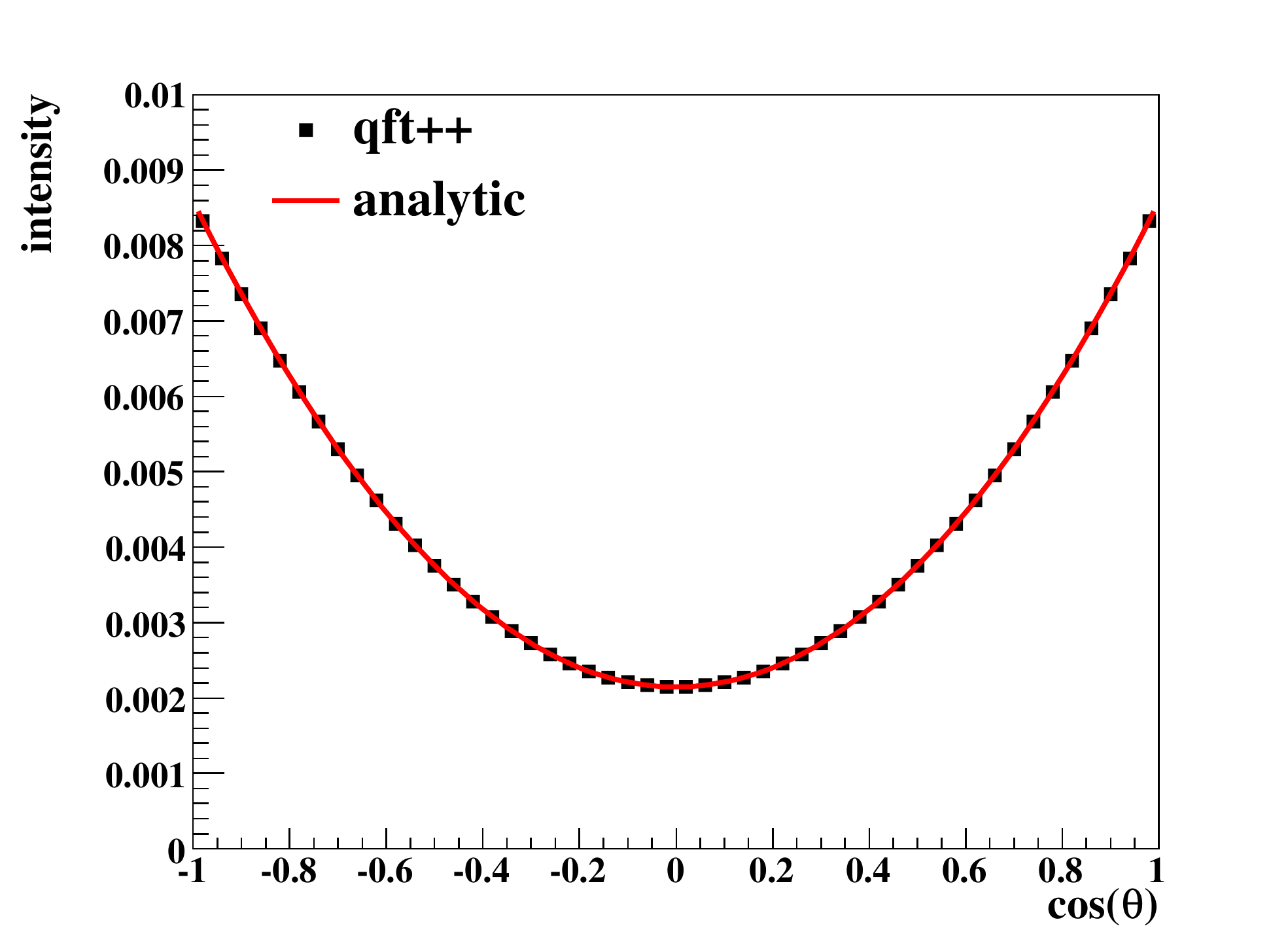}
  \caption[]{\label{fig:delta-dist}
    (Color Online)
    Intensity vs $\cos{\theta}$: Angular distribution 
    calculated for the reaction 
    ${\pi p \rightarrow \Delta(1232) \rightarrow \pi p}$ using 
    {\tt qft++} (black filled squares) compared with an analytic solution
    (red line). See text for details.
  }
  \end{center}
\end{figure}

Figure~\ref{fig:delta-dist} shows the angular distributions obtained by 
calculating (\ref{eq:delta-amp}) in the overall center-of-mass frame using
{\tt qft++} compared to the non-relativistic helicity formalism calculation
used to obtain (\ref{eq:delta-hel}). The result obtained using 
(\ref{eq:delta-hel}) was normalized to have the same integral as the 
covariant calculation. Clearly the two calculations give the same angular
distributions.

\subsection{\label{section:compton}Compton Scattering}

As a final example, the unpolarized Compton scattering cross section will be
calculated to leading order in $\alpha$.
The two well-known tree-level amplitudes, corresponding to $s$- and $u$-channel
electron exchange diagrams, are
\begin{subequations}
\label{eq:compton-amps}
  \begin{eqnarray}
    A_s = -e^2 \bar{u}(p',m_p')\gamma^{\mu}
   \epsilon^*_{\mu}(k',m_{\gamma}')
   \hspace{0.13\textwidth} \nonumber \\ \times
   \frac{\slashed{p}+\slashed{k} + w}{(p+k)^2 - w^2} 
   \gamma^{\nu}\epsilon_{\nu}(k,m_{\gamma}) u(p,m_p)
  \end{eqnarray} 
  \begin{eqnarray}
    A_u = -e^2 \bar{u}(p',m_p')\gamma^{\nu}
   \epsilon_{\nu}(k,m_{\gamma}) \hspace{0.13\textwidth}\nonumber \\ \times
   \frac{\slashed{p}-\slashed{k}' + w}{(p-k')^2 -w^2} 
   \gamma^{\mu}\epsilon^*_{\mu}(k',m_{\gamma}') u(p,m_p),
  \end{eqnarray}
\end{subequations} 
where $p(p')$ and $k(k')$ denote the initial(final) electron and photon momenta
respectively and $w$ is the mass of the electron. The full scattering 
amplitude, from which the cross section can be calculated,
is obtained by combining these two processes.

The calculation of this scattering amplitude is similar to that of the previous
example. First, the following variables must be declared:
\\ \\
\begin{boxedminipage}{0.475\textwidth}
\begin{verbatim}
DiracSpinor ui,uf; // e spinors
DiracSpinor u_ex; // exchanged e spinor
PolVector epsi,epsf; // photon pol.vecs
DiracGamma gamma; // gamma^mu
// s- and u-channel propagators
Matrix<complex<double> > prop_s;
Matrix<complex<double> > prop_u;
Vector4<double> pi,pf,ki,kf; // 4-momenta
\end{verbatim}
\end{boxedminipage}\\ \\ \\
A loop would then be performed over all values of $\cos{\theta}$ (scattering
angle in the lab frame) for which the cross section is to be calculated.
At each point, the propagators are obtained using the following code:
\\ \\
\begin{boxedminipage}{0.475\textwidth}
\begin{verbatim}
u_ex.SetP4(pi+ki,0.511);
prop_s = u_ex.Propagator();
u_ex.SetP4(pi-kf,0.511);
prop_u = u_ex.Propagator();
\end{verbatim}
\end{boxedminipage}\\ \\ \\
The scattering intensity is then obtained by looping over spin projections,
{\tt Spin m\_ef,m\_ei,m\_gf,m\_gi}, and calculating the amplitudes given in
(\ref{eq:compton-amps}) as
\\ \\
\begin{boxedminipage}{0.475\textwidth}
\begin{verbatim}
amp_s = Bar(uf(m_ef))*gamma
    *conj(epsf(m_gf))*prop_s*gamma
    *epsi(m_gi)*ui(m_ei);
amp_u = Bar(uf(m_ef))*gamma*epsi(m_gi)
    *prop_u*gamma*conj(epsf(m_gf))
    *ui(m_ei);
\end{verbatim}
\end{boxedminipage}\\ \\ \\
To get the cross sections the appropriate scale factors ($\alpha^2$, phase 
space factors, etc.) must then be applied.

\begin{figure}
  \begin{center}
  \includegraphics[width=0.5\textwidth]{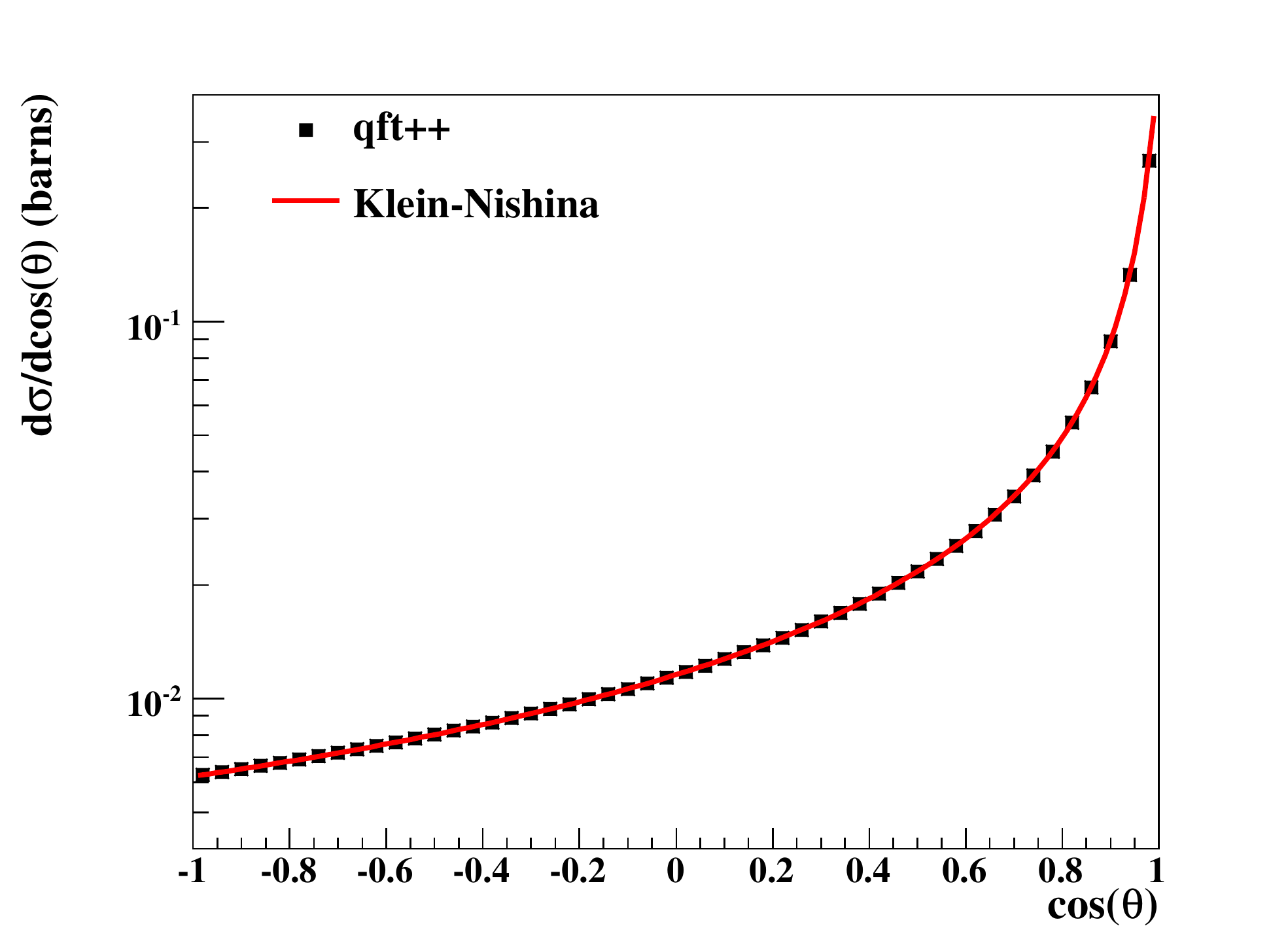}
  \caption[]{\label{fig:compton}
    (Color Online)
    $\frac{d\sigma}{d\cos{\theta}}$(barns) vs $\cos{\theta}$: 
    Compton scattering differential cross section for a 10~MeV incident photon
    calculated using
    {\tt qft++} (black filled squares) compared with the Klein-Nishina equation
    (red line). See text for details.
  }
  \end{center}
\end{figure}

Figure~\ref{fig:compton} shows the differential cross section for a 
10~MeV incident photon calculated using {\tt qft++} compared to the well-known
spin-averaged Klein-Nishina formula~\cite{cite:klein-nishina}.
Both methods give the same results.

\section{\label{section:examples:discuss}Discussion}

The examples in this paper were chosen because they can also easily be solved 
analytically, allowing for comparisons of the two calculations. 
In general, this is not the case; however, even for very complicated 
expressions involving high rank tensors or symmetrization, the {\tt qft++}
calculations are still very manageable 
(more complicated examples can be found online~\cite{cite:qft++website}). 
The code is also optimized for performance. 
The examples given in this paper run at 
$\sim3$~kHz, {\em i.e.} on a 2.5~GHz processor, one can compute results at 
about 3000 kinematic points per second.
This package makes it possible
for any physicist to compare and/or fit a theoretical model to his or her data
by coding up the expressions themselves.

It is clear from these examples how this package could be useful for a 
partial wave analysis, but it can also be used for other types of analyses.
For example, it can be used to perform numeric checks of analytic calculations.
The {\tt qft++} package can also be used for cases where analytic solutions
are desirable but not feasible. For example, consider the case where an
expansion of an expression in powers of some variable, $x$, is needed but an
analytic solution is not available. 
The {\tt qft++} package could be used to evaluate the expression at a large 
number of values of $x$. The coefficients in the expansion could then be 
extracted by fitting the {\tt qft++} results.

The {\tt qft++} package is available to all users from
{\tt http://www-meg.phys.cmu.edu/qft++}.
\begin{center}
\mbox{}\\\vspace{0.2in}
{\normalsize \textbf{Acknowledgments}}
\end{center}
\vspace{0.2in}
I would like to thank Matt Shepherd for providing a more efficient version 
of the {\tt TensorIndex} class.
This work was supported by grants from the United States Department of Energy
No. DE-FG02-87ER40315 and 
the National Science Foundation No. 0653316 through the 
``Physics at the Information Frontier'' program.



\begin{thebibliography}{99}
%
\bibitem{cite:feyncalc}
R. Mertig, M. Böhm and A. Denner.
Comput. Phys. Commun. {\bf 64} 345 (1991). 
%
\bibitem{cite:feynarts}
T. Hahn. Comput. Phys. Commun. {\bf 140}  418 (2001).
%
\bibitem{cite:grace}
J. Fujimoto {\em et al.} Comput. Phys. Commun. {\bf 153} 106 (2003).
%
\bibitem{cite:comphep}
E. Boos {\em et al.} Nucl.Instrum. Meth. {\bf A534} 250 (2004).
%
\bibitem{cite:williams-thesis}
M. Williams, Carnegie Mellon University Ph.D. Thesis, (2007).
%
\bibitem{cite:alexandrescu} 
A. Alexandrescu. {\em Modern C++ Design}. Addison-Wesley, 2001.
Using concepts discussed in Chapter 2.
%
\bibitem{cite:boost} 
J. J\"{a}rvi, J. Willcock and A. Lumsdaine. 
Following the {\tt enable\_if} family of templates contained in the {\tt BOOST}
libraries.
{\tt http://www.boost.org}. 
%
\bibitem{cite:rarita}
W. Rarita and J. Schwinger.
Phys. Rev. \textbf{60}, 61 (1941).
%
\bibitem{cite:zemach-form}
C. Zemach. 
Phys. Rev. \textbf{140}, B97 (1965).
%
\bibitem{cite:chung}
S.U. Chung. 
BNL preprint BNL-QGS-02-0900 (2004).
%
\bibitem{cite:anisovich}
A.V. Anisovich {\em et al}.
J. Phys. G \textbf{28}, 15-32 (2002).
%
\bibitem{cite:qft++website}
{\tt http://www-meg.phys.cmu.edu/qft++}
%
\bibitem{cite:zemach}
C. Zemach. 
Phys. Rev. B \textbf{133}, 1201-1220 (1964).
%
\bibitem{cite:klein-nishina}
O. Klein and Y. Nishina. Z. Physik \textbf{52}, 853 (1929).

\end{thebibliography}
\end{document}